\documentclass[twocolumn,floatfix,prl,aps,amsmath, amssymb]{revtex4-1}
\usepackage{graphicx}
\usepackage{amsmath}
\usepackage{amssymb}
\usepackage{color}
\usepackage{natbib}
\usepackage[pdftex,breaklinks,bookmarks=false,colorlinks,linkcolor=blue,citecolor=blue,urlcolor=blue]{hyperref}

\begin{document}
\title{Fibonacci Topological Superconductor}

\author{Yichen Hu and C. L. Kane}
\affiliation{Department of Physics and Astronomy, University of Pennsylvania, Philadelphia, PA 19104}

\begin{abstract}

 We introduce a model of interacting Majorana fermions that describes a superconducting phase with a topological order characterized by the Fibonacci topological field theory.   Our theory, which is based on a $SO(7)_1/(G_2)_1$ coset factorization, leads to a solvable one dimensional model that is extended to two dimensions using a network construction.   In addition to providing a description of the Fibonacci phase without parafermions, our theory predicts a closely related ``anti-Fibonacci" phase, whose topological order is characterized by the tricritical Ising model.   We show that Majorana fermions can split into a pair of Fibonacci anyons, and propose an interferometer that generalizes the $Z_2$ Majorana interferometer and directly probes the Fibonacci non-Abelian statistics.

\end{abstract}

\maketitle

Current interest in topological quantum phases is heightened by the proposal to use them for quantum information processing\cite{kitaev2006,nayak2008} and by prospects for realizing them in experimentally viable electronic systems.  There is growing evidence that the fractional quantum Hall (QH) state at filling $\nu=5/2$ is a non-Abelian state\cite{mooreread1991,Radu2008,dolev2008,Willett2009,Banerjee2017} with Ising topological order.   A simpler form of Ising order is predicted in topological superconductors (T-SC)\cite{readgreen2000,kitaev2001} and in SC proximity effect devices\cite{fukane2008,fukane2009,Lutchyn2010,Oreg2010,Qi2010}.   In these systems the Ising $\sigma$ particle is not dynamical, but is associated with domain walls or vortices that host gapless Majorana fermion modes.  Recent experiments have found promising evidence for Majorana fermions in 1D and 2D SC systems\cite{Mourik2012,NadjPerge2014,wang2017}.

Ising topological order is insufficient for universal quantum computation, but the richer Fibonacci topological order is sufficient\cite{Freedman2002}.   Fibonacci order arises in the $Z_3$ parafermion state introduced by Read and Rezayi\cite{readrezayi1999}, which is a candidate for the fractional QH state at $\nu = 12/5$.   Parafermions can also be realized by combining SC with the fractional QH effect\cite{Barkeshli2012,Lindner2012,Cheng2012,Clarke2013,Vaezi2013}.   This line of inquiry culminated in the tour de force works\cite{Mong2014,Vaezi2014} that showed a $\nu=2/3$ QH state, appropriately proximitized, could exhibit a Fibonacci phase.

In this paper we introduce a different formulation of the Fibonacci phase based on a model of interacting Majorana fermions.  Our starting point is a system of chiral Majorana edge states, which can in principle be realized in SC proximity effect structures.  We show that a particular four fermion interaction leads to an essentially exactly solvable model that realizes the Fibonacci phase.   In addition to providing a direct route to the Fibonacci phase {\it without parafermions}, our theory reveals a distinct but closely related ``anti-Fibonacci" state that is a kind of particle-hole conjugate to the Fibonacci state with a topological order that combines Ising and Fibonacci.    Our formulation also suggests a method for experimentally probing the Fibonacci state.  We introduce a generalization of the interferometer introduced earlier for Majorana states\cite{akhmerov2009,fukane2009b}, and argue that it provides a method for unambiguously detecting Fibonacci order.

The fact that interacting Majorana fermions can exhibit a Fibonacci phase is foreshadowed by Rahmani, et al. \cite{rahmani2015}(RZFA), who showed that a 1D Majorana chain with strong interactions can be tuned to the tricritical Ising (TCI) critical point.   The same critical point arises in  the 1D ``golden chain" model of coupled Fibonacci anyons\cite{feiguin2007}, as well as at interfaces connecting Ising and Fibonacci order in the QH effect\cite{Grosfeld2009}.  There is a sense in which the TCI point of the RZFA model is like a Fibonacci chain, but 
 it is not clear how to extend it to 2D.   Our theory provides a method for accomplishing that.  

Mong et al. \cite{Mong2014} formulated the Fibonacci phase using a ``trench" construction that began with 1D strips of $\nu=2/3$ QH states coupled along trenches in the presence of a SC.   A single trench mapped to the 3 state clock model, with a critical point described by the $Z_3$ parafermion conformal field theory (CFT).  The resulting 1D states were coupled to create a gapped 2D phase.   
This is similar to the coupled wire construction\cite{kml2002} for the Read Rezayi state introduced in Ref. \onlinecite{teokane2014}, but differs in an important way.  That model was based on the coset construction\cite{Goddard1985,Fradkin1999,byb}, which allows a simple CFT ($[SU(2)_1]^3$ with central charge $c=3$) to be factored into less trivial CFTs ($SU(2)_3 + SU(2)_1^3/SU(2)_3$ with $c= 9/5 + 6/5$).   This exact factorization identifies a solvable coupled wire Hamiltonian, where counter-propagating modes of the two factors pair up differently, resulting in a non-trivial unpaired chiral edge mode\cite{teokane2014,Teo2016}.

The construction in this paper is based on the coset $SO(7)_1/(G_2)_1$\cite{vafa1994}.   $SO(7)_1$ describes $7$ free chiral Majorana modes with $c=7/2$.  $G_2$ is a Lie group that sits inside $SO(7)$.   $(G_2)_1$, with $c=14/5$, is the Fibonacci CFT\cite{bonderson2007,Mong2014}.   The quotient is a CFT with 
\begin{equation}
c=7/2 - 14/5 = 7/10, 
\end{equation}
which can be identified with the TCI model.   Thus, the edge states of a non-interacting T-SC with Chern number $n=7$ factor into a $(G_2)_1$ Fibonacci (FIB) sector and a $SO(7)_1/(G_2)_1$ TCI sector.   In the following we will design an interaction that separates the factors and leads to $2D$ topological phases with either $c=14/5$ (Fibonacci) or $c=7/10$ (anti-Fibonacci) edge states.

We begin with some facts about $G_2$, which is well known in mathematical physics\cite{vafa1994,ketov1996}.
$G_2$ is the simplest exceptional Lie group. Its relation to $SO(7)$ involves the mathematics of the octonion division algebra\cite{baez2002}.
 An octonion is specified by 8 real numbers:  $q = q_0  + \sum_{a=1}^7 q_a e_a$, where $e_a$ are 7 square roots of $-1$ that satisfy the non-associative multiplication rule
\begin{equation}
e_a e_b = - \delta_{ab} + C_{abc} e_c.  
\label{octonion multiplication}
\end{equation}
$C_{abc}$ is a totally antisymmetric tensor.  It is not unique, but
can be chosen to satisfy\cite{baez2002}
\begin{equation}
C_{a+1b+1c+1} = C_{abc}, \quad C_{124}=1,
\label{124}
\end{equation}
where the indices are defined mod 7.  Eq. \ref{124} along with antisymmetry specifies all the non-zero elements of $C_{abc}$.   $e_a$ define a set of 7 unit vectors that transform under $SO(7)$.   However, not all $SO(7)$ rotations preserve (\ref{octonion multiplication}).   $G_2$ is the automorphism group of the octonions: the subgroup of $SO(7)$ that preserves $C_{abc}$.

The 21 generators of $SO(7)$ can be represented by $7 \times 7$ skew symmetric matrices $T^{m,n}$ of the form $T^{m,n}_{ab} = i(\delta_{ma}\delta_{nb}-\delta_{mb}\delta_{na})$.   There are 14 combinations that preserve $C_{abc}$, which can be written\cite{ketov1996}
\begin{equation}
M^A = \left\{\begin{array}{cl} \frac{T^{A,A+2} - T^{A+1,A+5}}{\sqrt{2}} & 1\le A\le 7  \\
 \frac{T^{A,A+2}+T^{A+1,A+5}-2 T^{A+3,A+4}}{\sqrt{6}} & 8 \le A \le 14.\end{array}\right.
\end{equation}
These matrices are normalized by ${\rm Tr}[M^A M^B] = 2\delta_{AB}$ and represent the generators of $G_2$ in the 7D fundamental representation, analogous to the Pauli matrices of $SU(2)$.
In what follows, it will be useful to express the quadratic Casimir operator as 
\begin{equation}
\sum_A M^A_{ab} M^A_{cd} = \frac{2}{3}(\delta_{ad}\delta_{bc}-\delta_{ac}\delta_{bd}) - \frac{1}{3} *C_{abcd}
\label{casimir}
\end{equation}
where $*C_{abcd} = \epsilon_{abcdefg}C_{efg}/6$ is the dual of $C_{abc}$ whose non-zero elements follow from $*C_{3567}=-1$, as in (\ref{124}).

\begin{figure}
\includegraphics[width=3.4in]{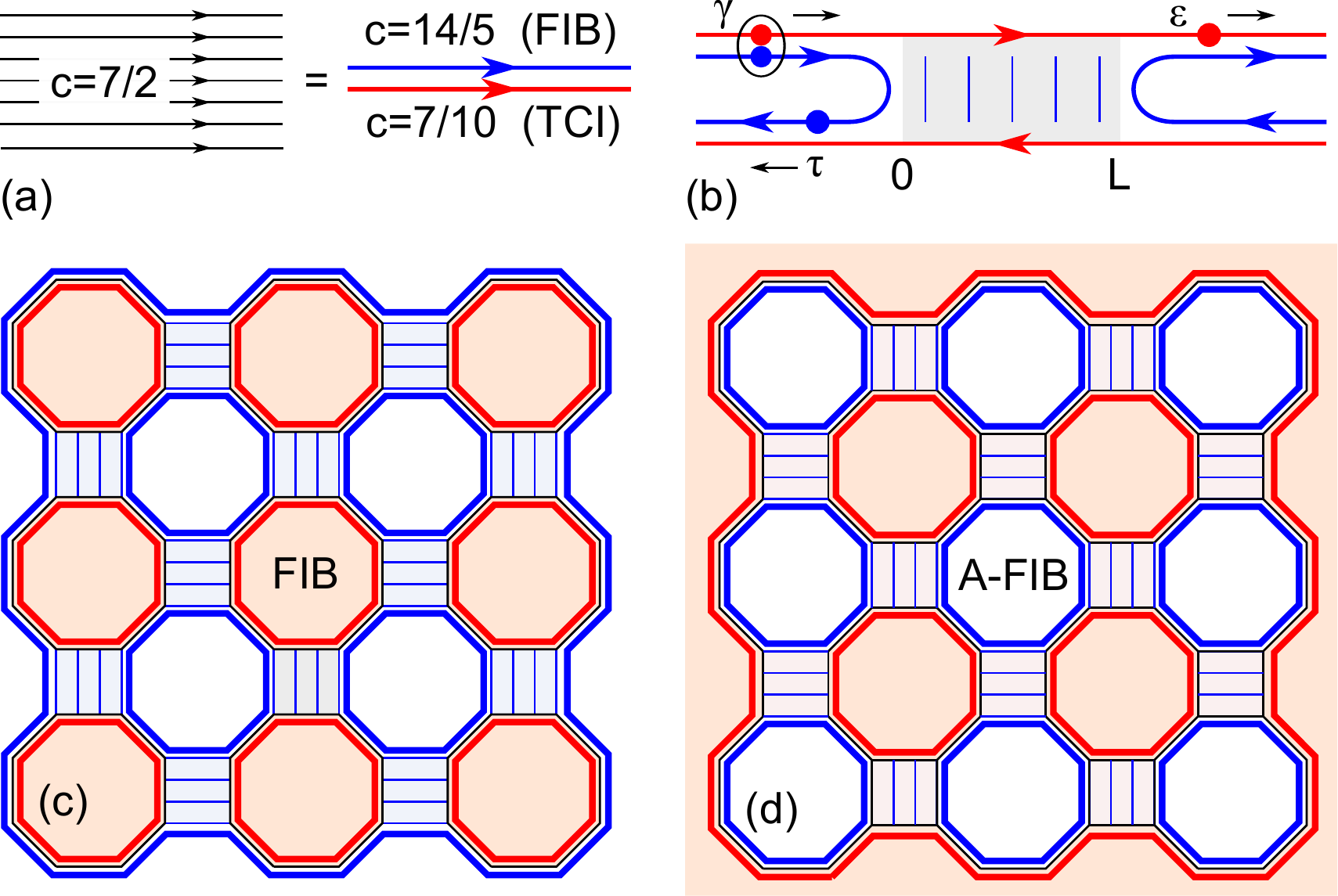}
\caption{(a) 7 chiral Majorana edge modes factor into FIB and TCI sectors with $c=14/5+7/10=7/2$.  (b) A 1D non-chiral system with interaction $\lambda\sum_A J^A_R J^A_L$ transmits the TCI sector, but reflects the FIB sector.  The bottom panels show network constructions for the Fibonacci phase (c) and the anti-Fibonacci phase (d). }
\label{Fig1}
\end{figure}

We now consider the coset factorization of a 1D system of 7 free chiral Majorana fermions described by
\begin{equation}
H_0 = -\frac{i v}{2}\sum_{a=1}^7 \gamma_a \partial_x \gamma_a.
\end{equation}
We adopt a Hamiltonian formalism\footnote{The Hamiltonian density is equivalent to the CFT energy momentum tensor on a cylinder: $H_0 = v T_{\rm cyl}/2\pi$.}
with Majorana operators satisfying $\{\gamma_a(x),\gamma_b(x')\}=\delta(x-x')\delta_{ab}$. 
 $H_0$ 
describes a $SO(7)_1$ Wess Zumino Witten (WZW) model with $c=7/2$.  The coset construction allows this to be written $H_0 = H_{\rm FIB} + H_{\rm TCI}$.  The FIB sector is expressed in terms of $(G_2)_1$ currents in Sugawara form
\cite{byb}\footnote{In addition to \cite{Note1}, $J^A$ differs in normalization from the WZW current defined in Ref. \cite{byb}, which is $2\pi J^A$ },
\begin{equation}
H_{\rm FIB} = \sum_A \frac{\pi v J^A J^A}{k+g}, \quad J^A = \sum_{ab} \frac{1}{2}M^A_{ab} \gamma_a \gamma_b,
\label{J^A}
\end{equation}
with $k=1$, $g=4$.  
Using (\ref{casimir}), the operator product gives
\begin{align}
&H_{\rm FIB}  = -\frac{2 i v}{5}\sum_a\gamma_a\partial_x\gamma_a - \frac{\pi v}{60}\sum_{abcd} *C_{abcd}\gamma_a\gamma_b\gamma_c\gamma_d, \nonumber\\
 &H_{\rm TCI} =  -\frac{i v}{10}\sum_a \gamma_a\partial_x\gamma_a +\frac{\pi v}{60} \sum_{abcd} *C_{abcd}\gamma_a\gamma_b\gamma_c\gamma_d.
\end{align}
The correlator of $H_{\alpha={\rm FIB,TCI}}$ is $\langle H_\alpha(x)H_\beta(x')\rangle = v^2\delta_{\alpha\beta}c_\alpha/8\pi^2(x-x')^4$, with $c_{\rm FIB} = 14/5$ and $c_{\rm TCI}=7/10$
\footnote{Note that \unexpanded{$\langle\gamma_a(x)\gamma_b(0)\rangle = \delta_{ab}/2\pi i x$}  and
$\sum *C^2_{abcd}=168$.}.
This shows that $H_0$ decouples into two independent sectors, as depicted in Fig.  \ref{Fig1}a.

$H_{\rm FIB}$ describes a $(G_2)_1$ WZW model, with two primary fields $1$, $\tau$ of dimension $h= 0$, $2/5$.   $\tau$ transforms under the 7D representation of $G_2$ and obeys the Fibonacci fusion algebra $\tau\times\tau=1+\tau$.   $H_{\rm TCI}$  describes the $M(5,4)$ minimal CFT with 6 primary fields $1$, $\epsilon$, $\epsilon'$, $\epsilon''$, $\sigma$, $\sigma'$, with $h=0$, $1/10$, $3/5$, $3/2$, $3/80$, $7/16$\cite{byb}.   The Majorana fermion operator $\gamma_a$ factors into the product 
\begin{equation}
\gamma_a = \tau_a \times \epsilon
\end{equation}
with $h=2/5+1/10=1/2$.   The 21 bilinears $i\gamma_a\gamma_b$ decompose into 14 $J^A$'s, along with 7 operators $\tau_a \times \epsilon'$ with $h=2/5+3/5=1$.    $J^A$ act only in the FIB sector: $[J^A,H_{\rm TCI}]=0$.  The trilinear combination $C_{abc}\gamma_a\gamma_b\gamma_c$ is $\epsilon''$ with $h=3/2$ and acts only in the TCI sector. 

We now introduce a 1D model of 7 non-chiral Majorana fermions $\gamma_{aR/L}$with an interaction that gaps the FIB sector, leaving the TCI sector gapless.  Consider
\begin{equation}
H = -\frac{i v}{2}\sum_a (\gamma_{aR}\partial_x\gamma_{aR} - \gamma_{aL}\partial_x\gamma_{aL}) + \lambda\sum_A J^A_R J^A_L,
\label{hlambda}
\end{equation}
where $J^A_{R/L}$ are given in (\ref{J^A}).  The $\lambda$ term commutes with $H_{\rm TCI}$, so it operates only in the FIB sector.  A perturbative renormalization group analysis gives $d\lambda/d\ell = - 2 \lambda^2/\pi v$, so $\lambda<0$ is marginally relevant.   When $\lambda$ flows to strong coupling it is natural to expect that it leads to a gap $\Delta \propto e^{-\pi v/2|\lambda|}$ in the FIB sector and a gapless TCI critical point.  This is similar to the RZFA model, 
except the $G_2$ symmetry locates the critical point exactly.   

The exact factorization allows the two sectors to be separated.
Consider the 1D system in Fig. \ref{Fig1}b, with $\lambda(x)\ne 0$ for $0<x<L$.  Provided $L \gg \xi=v/\Delta$, the gap in the FIB sector leads to an exponential suppression of transmission.  The FIB sector will be perfectly reflected, while the TCI sector will be perfectly transmitted.   Interestingly, this means an incident Majorana fermion $\gamma_a$ {\it splits}, with $\tau_a$ reflected and $\epsilon$ transmitted.   This forms the basis for the interferometer to be discussed below.  

We wish to use (\ref{hlambda}) to construct a 2D gapped topological phase.   One approach is to adapt the coupled wire model\cite{kml2002}.   This requires coupling right movers of the TCI sector on wire $i$ to left movers of the TCI sector on wire $i+1$.   If this gaps the TCI sector, then we will have a 2D gapped phase with TCI edge states.   This is problematic, however, because the simplest tunneling term that can be built from local operators and does not couple to the gapped FIB sector is the trilinear $C_{abc}\gamma_a\gamma_b\gamma_c$.   The resulting tunneling term $u \epsilon''_{iR}\epsilon''_{i+1L}$, with dimension 3, is perturbatively irrelevant.   This does not preclude the possibility of a gapped phase for large $u$, but a non-perturbative analysis would be necessary to establish it.   
Fortunately, however, the exact factorization of the coset model allows for an alternative {\it network construction}, inspired by the Chalker Coddington model\cite{Chalker1988}.

Fig. \ref{Fig1}c shows a network of $n=7$ T-SC islands in which each island has 7 chiral Majorana modes.   In the absence of coupling the Majorana modes are localized on each island, so the system is a trivial SC.   If the islands are strongly coupled by single particle tunneling they will merge, and the system is a $n=7$ T-SC.   In the absence of interactions, the transition between these phases will have 7 gapless $2+1D$ Majorana modes.   For strong interactions intermediate topological phases can arise.   We turn off the single particle tunneling and couple the neighboring islands with the interaction term in (\ref{hlambda}).   Provided the contact length $L\gg \xi$, the excitations in the FIB sector will be reflected from the contact, which means they are transmitted to the next island.   Excitations in the TCI sector, however, are transmitted by the contact, so they remain localized on the same island.    From Fig. \ref{Fig1}c, it can be seen that both the TCI and the FIB sectors are localized in the interior of the network.   The TCI states are localized on the islands, while the FIB states are localized on the dual lattice of voids between the islands.   Since all bulk states are localized in finite, lattice scale regions, there will be a bulk excitation gap.   The perimeter of the network, however has a gapless FIB edge state with $c=14/5$.   
We emphasize that though fine tuning is required to achieve the exactly solvable Hamiltonian (\ref{hlambda}), the tuning does {\it not} need to be perfect.  This gapped Fibonacci {\it phase} will be robust to finite single particle tunneling and other interactions.

Fig. \ref{Fig1}d shows a similar network that is surrounded by a $n=7$ chiral Majorana edge state.   This leads to a distinct phase that also has a bulk gap, but has TCI edge states with $c=7/10$.   This state can be viewed as a Fibonacci phase sitting inside a $n=7$ T-SC, with $c=7/2-14/5$.   We call this the ``anti-Fibonacci" in analogy with the ``anti-pfaffian"
\cite{levin2007,lee2007}, which is the pfaffian sitting inside a $\nu=1$ QH state.     The anti-Fibonacci has a topological order associated with the TCI CFT.   However, the 6 TCI quasiparticles can also be understood as a combination of $1$, $\tau$ Fibonacci quasiparticles with the $1$, $\psi$, $\sigma_i$ Ising quasiparticles.  The TCI fusion rules\cite{byb} of the quasiparticles identified in Table \ref{tcifields} are reproduced by the simpler Fibonacci and Ising fusion rules (e.g. $\sigma_i\times\sigma_i = 1 + \psi$).  Similar fusion rule decompositions have been identified for other theories\cite{bonderson2007,Grosfeld2009}.   As in the T-SC $\sigma$ and $\sigma'$ are not dynamical quasiparticles, but they will be associated with $h/2e$ vortices in the SC.   Depending on the energetics, a SC vortex in the anti-Fibonacci phase will bind either a $\sigma$ or $\sigma'$.   If it is $\sigma$, then the vortex binds a Fibonacci anyon.  Likewise in the Fibonacci phase, a vortex could bind $1$ or $\tau$\cite{Mong2014}.

\begin{table}
\begin{ruledtabular}
\begin{tabular}{@{\qquad}  c @{\qquad} |@{\qquad} c @{\qquad}  @{\qquad} c @{\qquad}   @{\qquad} l @{\qquad}}
 & $1$ & $\psi$ & $\sigma_i$ \\
 \hline
 $1$  & $1$ &$\epsilon''$ & $\sigma'$  \\
  $\tau$ & $\epsilon'$ & $\epsilon$ & $\sigma$\\
\end{tabular}
\end{ruledtabular}
\caption{The 6 quasiparticles of the TCI model can be identified with combinations of Ising and Fibonacci quasiparticles.}
\label{tcifields}
\end{table}

The above considerations suggest a possible route towards realizing the Fibonacci phase is to start with a system close to a multi-component T-SC - trivial SC transition.   This could be achieved by introducing SC via the proximity effect into a 2D electron gas in the vicinity of a quantum Hall plateau transition with degenerate Landau levels.   Progress in this direction has recently been reported in a quantum anomalous Hall insulator coupled to a SC, where a plateau observed in the two terminal conductance was attributed to T-SC\cite{wang2017}.   Another promising venue is graphene, which has a four-fold degenerate zeroth Landau level.   Coexistence of SC with the quantum Hall effect in these systems appears feasible\cite{amet2016,yacoby2016}.

If the Fibonacci and/or the anti-Fibonacci T-SC can be realized, then it will be important to develop experimental protocols for probing them.   One approach is to measure the thermal Hall conductance, which directly probes the central charge $c$ of the edge states:  $\kappa_{xy} = c \pi^2 T k_B^2/3h$.   This has proven to be a powerful method for identifying the topological order of QH states\cite{kanefisher1997,Banerjee2016,Banerjee2017}, but it does not directly probe the non-Abelian quasiparticle statistics.   In the QH effect, Fabry Perot\cite{Stern2006,Bonderson2006a,Bonderson2006} and Mach Zehnder\cite{Feldman2006,Law2008} interferometers have been proposed for this purpose.   Here we introduce a distinct interferometer that generalizes the Majorana fermion interferometer\cite{akhmerov2009,fukane2009b}.

\begin{figure}
\includegraphics[width=3.3in]{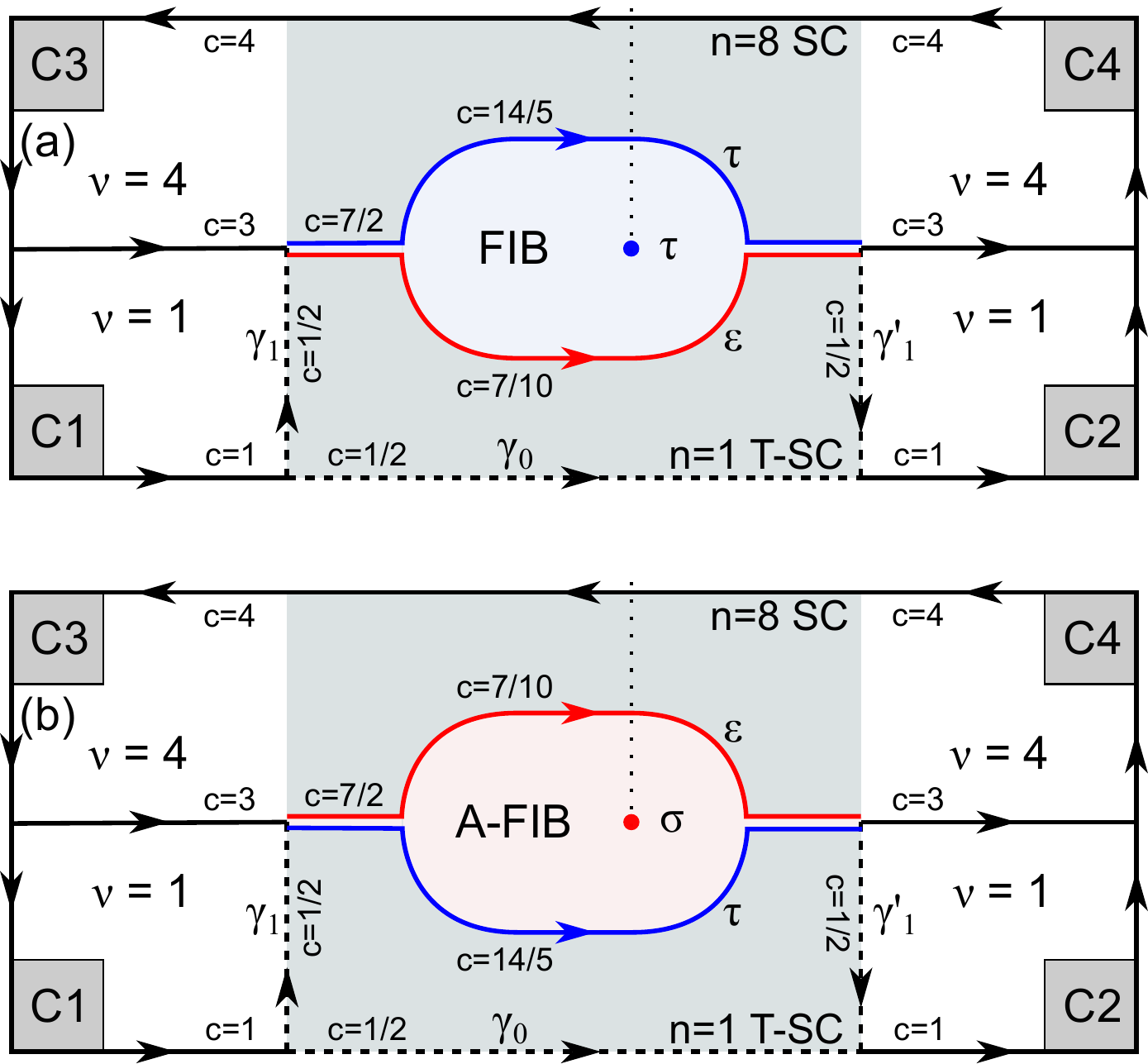}
\caption{A Fibonacci interferometer in a Hall bar with Ohmic contacts C1-4 and SC in the shaded region.  Dirac (Majorana) edge states are indicated by solid (dashed) lines.   The $c=7/2$ edge splits into FIB and TCI edges around the Fibonacci (a) or anti-Fibonacci (b) island.   A quasiparticle adds a branch cut (dotted line) that modifies transmission from C1 to C2.  }
\label{Fig2}
\end{figure}

Fig. \ref{Fig2} shows a Hall bar with 4 Ohmic contacts (C1-4) where the electron density is adjusted so that adjacent regions have  QH filling factors $\nu=1$ and $\nu=4$.    The middle is coupled to a SC that leads to a $n=1$ T-SC region and a trivial $n=8$ SC region.    We assume that at the boundary between the $n=1$ and $n=8$ SCs there is an island of either Fibonacci (Fig. \ref{Fig2}a) or anti-Fibonacci (Fig. \ref{Fig2}b).    This leads to the pattern of edge states shown.      

Suppose contact $C1$ is at voltage $V_1$, and that the SC and the other 3 contacts are grounded.   We use a Landauer-B\"uttiker formalism\cite{Buttiker1988} to compute the current in C2, given by $I_2 = v_F \langle\psi_{\rm in}^\dagger\psi_{\rm in} - \psi_{\rm out}^\dagger\psi_{\rm out}\rangle$, where $\psi_{\rm in(out)}$ describe the $\nu=1$ chiral fermions entering (leaving) C2.   $V_1$ only affects $\psi_{\rm in}$, which in the SC decomposes into $\gamma_0 + i \gamma_1'$\cite{akhmerov2009,fukane2009b}.  Thus $I_2 \propto \langle i\gamma_0 \gamma'_1\rangle$.
$\gamma_0$ comes directly from C1, but $\gamma'_1$ comes from the region where $\tau$ and $\epsilon$ split and then recombine.  First suppose there are no quasiparticles on the island.  $\gamma'_1$ will be a linear combination $\sum_{j=1}^7 t_{1j} \gamma_j$ of the incident Majorana modes, where $t_{ij}$ is a real orthogonal scattering matrix and $\gamma_{2-7}$ are associated with the $c=3$ edge.   Ignoring the contributions from the grounded contact C3, $i \gamma_0\gamma'_1 = t_{11} i\gamma_0\gamma_1$.  This relates $I_2$ to the current coming out of C1, $I_2 = t_{11} (e^2/h) V_1$.     

Quasiparticles localized on the island will modify this result.  The transmitted particles will encounter a branch cut due to non-Abelian statistics that can modify the state of the localized quasiparticle.  Provided the local Hamiltonian near the edge is not modified by the presence of the extra quasiparticle, this will be purely of topological origin.   The expectation value of the current will only be non-zero if the localized quasiparticle returns to its original state.   The probability amplitude that anyon $a$ returns to its original state when circled by anyon $b$ is given by the {\it monodromy matrix}\cite{Bonderson2006} $M_{ab} = S_{ab}S_{11}/S_{a1}S_{b1}$, which depends the topological data in the modular S-matrix $S_{ab}$.   We therefore predict
\begin{equation}
I_2 = \frac{e^2}{h} t_{11} M_{ab} V_1,
\end{equation}
where $a$ and $b$ are labels for the transmitted and localized quasiparticles.    Provided quasiparticles can be introduced to the island without modifying $t_{11}$, (which depends on the local Hamiltonian near the edges) the ratios of the conductances for different localized quasiparticles will be {\it universal} (note $M_{a1}=1$).    Other proposed interferometric measurements of Fibonacci statistics have challenges similar to controlling $t_{11}$\cite{Bonderson2006,bonderson2007}.   A possible (albeit more complicated) way to overcome that is to include a contact inside the island that allows quasiparticles to come and go, leading to telegraph noise\cite{kane2003}.

For the FIB phase, where the transmitted quasiparticle is $\tau$ the universal ratio is determined by
\begin{equation}
M^{\rm FIB}_{\tau\tau} = -1/\varphi^2,
\end{equation}
where $\varphi = (1+\sqrt{5})/2$ is the golden mean.   In the A-FIB phase, the ratios are determined by $M^{\rm TCI}_{\epsilon b}$ for $b=1, \epsilon, \epsilon', \epsilon'', \sigma, \sigma'$.   These can be evaluated from the $6\times 6$ TCI S-matrix\cite{rahmani2015}.   However, the same results are obtained by treating the A-FIB as the FIB sitting inside Ising.   Then, $M^{\rm TCI}_{\epsilon b} = M^{\rm I}_{\psi b_i} M^{\rm FIB}_{\tau b_f}$, where $b_{i(f)}$ are the Ising (Fibonacci) decomposition of particle $b$ from Table \ref{tcifields}.  The non-trivial Ising term is $M^{\rm I}_{\psi \sigma_i} = -1$  (which is probed in the Majorana interferometer).   In the A-FIB state, if a vortex binds $\sigma$, the extra quasiparticle can be controlled with a magnetic flux, and $M^{\rm TCI}_{\epsilon\sigma}=+1/\varphi^2$.

In this paper we have introduced a theory of the Fibonacci phase based on Majorana fermions near a multicomponent topological critical point with strong interactions.   While this phase has the same topological structure as the parafermion based Fibonacci states, our theory clarifies the relation between the Fibonacci and ant-Fibonacci phases and shows the way in which Majorana fermions can fractionalize into Fibonacci anyons.
It also points to a promising direction in the broader problem of searching for exotic topological phases in strongly interacting systems with massless single-particle Dirac or Majorana fermions.

\acknowledgments

It is a pleasure to thank Jonathan Heckman, Abhay Pasupathy, Ady Stern and Jeffrey Teo for helpful discussions.   This work was supported by a Simons Investigator grant from the Simons Foundation.

%
%

%

\end{document}